\documentclass{svproc}
\usepackage{url}

\usepackage{graphicx}
\usepackage{hyperref}
\usepackage{url}

\begin{document}
\mainmatter              

\title{Event-by-event charge separation in Pb-Pb collisions at $\sqrt{s_{\rm NN}}$ = 2.76 TeV with ALICE at the LHC}

\titlerunning{Charge Separation}  
%
\author{Sonia Parmar (for the ALICE Collaboration)} 
\institute{Panjab University, Chandigarh - 160014, India\\
\email{sonia.parmar@cern.ch}} 
\maketitle              

\vspace{-0.3 in}
\begin{abstract}
Relativistic heavy-ion collisions provide a unique opportunity to search for parity violation in non-central collisions. This could lead to charge separation perpendicular to the reaction plane. An event-by-event measurement of charge separation effect in Pb-Pb collisions at $\sqrt{s_{\rm NN}}$ = 2.76 TeV using Sliding Dumbbell Method (SDM) is discussed in this article. 
\end{abstract}
\vspace{-0.4 in}
\section{Introduction}
\begin{figure} [h]         
\begin{center}                                                                                                
\includegraphics[scale=.195]{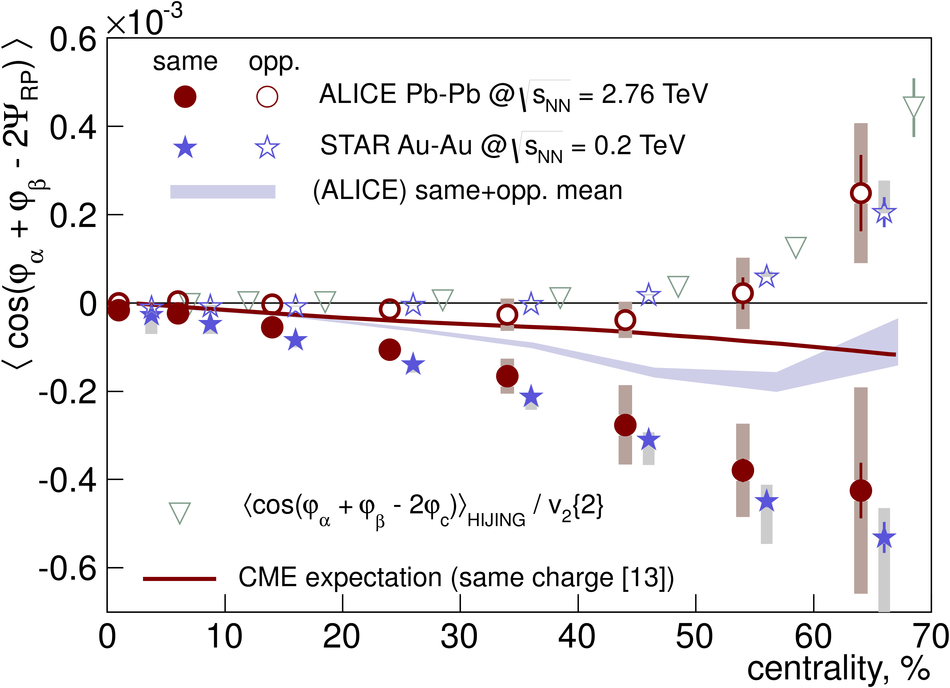}
\vspace{-0.19 in }
\caption{\label{fig1} Centrality dependence of the correlator for same sign charged pairs  (red colour) and opposite sign charged pairs  (blue colour) \cite{ALICEprl}.}
\vspace{-0.5 in }
\end{center}
\end{figure}

\vspace{-0.1 in}
ALICE, A Large Ion Collider Experiment, is specially optimized to study the properties of the deconfined state of quarks and gluons known as the ``Quark Gluon Plasma (QGP)''. In non-central heavy-ion collisions, strong parity violation coupled with a strong magnetic field leads to a charge separation along the system's angular momentum direction resulting in the ``Chiral Magnetic Effect (CME)''~\cite{CME}. This effect has been studied by the STAR experiment at RHIC  for different beam energies $\sqrt{s_{\rm NN}}$= 7.7 - 200 GeV~\cite{STARprl} and by the ALICE at LHC at $\sqrt{s_{\rm NN}}$ = 2.76 TeV~\cite{ALICEprl}. In view of this, Voloshin~\cite{voloshin} proposed a multi-particle correlator $\langle cos(\phi_\alpha + \phi_\beta - 2 \Psi_{RP})\rangle$ to observe CME, where $\phi_\alpha$, $\phi_\beta$ are the azimuthal angles of the particles $\alpha$, $\beta$, respectively and $\psi_{RP}$ is the reaction plane angle. The ALICE collaboration investigated the charge separation effect by studying the centrality dependence of this multi-particle correlator for the different charge combinations such as opposite sign charge pairs (+ -) and same sign charge pairs (++, - -), as shown in Fig.~\ref{fig1}~\cite{ALICEprl}. This indicates that the correlation for same sign charge pairs increases as one moves from the central to peripheral collisions. Fig.~\ref{fig1}~\cite{ALICEprl} also displays the STAR data points, which exhibits the similar trend.

For this analysis, the ALICE time projection chamber (TPC) covering the pseudorapidity region, $\eta < |0.9|$ is used to reconstruct the charged particle tracks and the VZERO detector is used to measure the collision centrality. The accepted events are divided into various centrality classes from 20-30$\%$ to 60-70$\%$. The analysis is performed in pseudorapidity coverage $\eta < |0.8|$ and 0.2 $< p_{\rm T} <$ 5 GeV/c. 
\vspace{-0.15 in }
\section{Sliding Dumbbell Method}
\vspace{-0.11 in}
To investigate the localized charge separation on event-by-event basis, the SDM is used, which is similar to the Sliding Window Method~\cite{SWM} used for charged-neutral fluctuations in Pb-Pb collisions at 158 A GeV at SPS~\cite{sps}. In this method, the observable $Db_{+-}$ is measured and defined as:

\vspace{-0.34 in }
\begin{equation}
{Db_{+-}} = {\frac{N^L_{+}}{(N^L_{+} + N^L_{-})}~ + ~\frac{N^R_{-}}{(N^R_{+} + N^R_{-})} ~,}
\vspace{-0.05 in }
\end{equation}

where, N$^L_{+}$ and N$^L_{-}$ are the numbers of positively and negatively charged particles on the left side of the dumbbell respectively, whereas N$^R_{+}$ and N$^R_{-}$ are the numbers of positively and negatively charged particles on the right side of the dumbbell respectively. The whole azimuthal plane is scanned by sliding the dumbbell of ${\Delta \phi}$ =$90^{\circ}$ in steps of ${\delta \phi}$ =$1^{\circ}$ and calculating the fraction $Db_{+-}$ in each event to extract the maximum value of $Db_{+-}$. The $Db_{+-}$ can also be calculated for different centrality intervals ranging from 20-30$\%$ to 60-70$\%$.

\vspace{-0.15 in}
\section{Results and Discussion}
\vspace{-0.1 in}

A sample of about 1 million minimum bias events of Pb-Pb collision data at $\sqrt{s_{\rm NN}}$ = 2.76 TeV were analysed. The $Db_{+-}$ distributions are obtained for different centralities using the SDM and also obtained by placing the dumbbell at randomly chosen azimuth in each event for the similar centrality intervals. $Db_{+-}$ distribution obtained from randomly placed dumbbells shows a gaussian distribution, which peaks $\sim$ 1 but the one obtained using SDM shifts towards higher $Db_{+-}$ values and this shift is more for semi-central events as compared to those for central events. Some of the semi-central events have been observed with much larger $Db_{+-}$ values, which indicates that the positive charged particles are on one side of the dumbbell and the negative charged particles are on the other side of the dumbbell. The percentage of such events increases with decreasing collision centrality. Further studies on Monte Carlo Generators without the CME (e.g. HIJING) are needed to investigate any biases of this method.
\vspace{-0.15 in }

\end{document}